\documentclass[
	a4paper,
	pagesize,
	pdftex,
	12pt,
	english,
	fleqn,
	final,
    captions=tableabove, %
	]{scrartcl}
\usepackage{ucs}
\usepackage[utf8x]{inputenc} %
\usepackage{fixltx2e} %
\usepackage[T1]{fontenc} %
\usepackage[english]{babel}
\usepackage{lmodern} %
\usepackage[numbers]{natbib}
\usepackage[pdfusetitle,pdftex,colorlinks,allcolors=blue,hyperfootnotes=false]{hyperref}
\usepackage{setspace,graphicx,tikz,tabularx} %
\usepackage[draft=false,babel,tracking=true,kerning=true,spacing=true]{microtype} %
\usepackage{subcaption}
\usepackage{csquotes}
\usepackage{tikz}
\usetikzlibrary{graphs,graphs.standard,topaths}
\usepackage[colorinlistoftodos, textsize=tiny]{todonotes}
\graphicspath{{figures/}}
\usepackage[nolist]{acronym}

\usepackage{amsmath}
\usepackage{amssymb}
\usepackage{mathtools}
\DeclareMathOperator{\V}{\mathbf{V}}
\DeclareMathOperator{\Q}{\mathbf{Q}}
\DeclarePairedDelimiter\set{\{}{\}}
\DeclarePairedDelimiter\len{\lvert}{\rvert}

\newcommand{\code}[1]{\texttt{#1}}

\begin{acronym}[acronyms]
    \acro{fbas}[FBAS]{Federated Byzantine Agreement System}
    \acro{mcp}[MCP]{MobileCoin Consensus Protocol}
    \acro{scp}[SCP]{Stellar Consensus Protocol}
\end{acronym}

\title{Crawling the MobileCoin Quorum System}
\author{Charmaine Ndolo, Sebastian Henningsen, Martin Florian\\
\small Humboldt University of Berlin / Weizenbaum Institute%
}
\date{}

\begin{document}

\maketitle

\begin{abstract}%
    We continuously crawl the young MobileCoin network,
    uncovering the quorum configurations of core nodes and the
    quorum system
    resulting from these configurations.
    This report discusses our crawl methodology, encountered challenges,
    and our current empirical results.
    We find that the MobileCoin quorum system currently comprises of 7 organisations controlling a total of 10 validator nodes.
    Current quorum set configurations prioritise safety over liveness.
    At the time of writing,
    one of the involved organisations is technically able to block the approval of new blocks,
    as is the case for one of the (two) ISPs employed by crawled nodes.
\end{abstract}

\acused{mcp}
\acused{scp}
\section{Introduction}
\emph{MobileCoin}\footnote{\url{https://www.mobilecoin.com/}} is a young peer-to-peer payments network intended for use on mobile devices~\cite{mobilecoin2017},
most prominently via integration into the popular messaging app \emph{Signal}\footnote{\url{https://signal.org/blog/update-on-beta-testing-payments/}}.
New transactions are added to a distributed public ledger via the \emph{\acl{mcp}} (\acs{mcp}), an implementation of the \emph{\acl{scp}} (\acs{scp}) \cite{mazieres2015stellar}.
Like \ac{scp}, \ac{mcp} functions on the basis of a \emph{\acl{fbas}} (\acs{fbas}):
each participating node can individually choose which groups of validators it requires agreement from.
Like Stellar~\cite{lokhava2019stellar_payments},
MobileCoin can therefore be placed on the middle ground of the permissioned--permissionless spectrum~\cite{florian2021sum}.

As a distinctive feature, MobileCoin uses \emph{Intel Software Guard eXtensions} (Intel SGX) to provide \enquote{defense-in-depth improvements to privacy and trust}\footnote{\url{https://github.com/mobilecoinfoundation/mobilecoin\#consensus/}}.
Hence, node operators who wish to join the network as validators need hardware with Intel SGX support.

The aim of the presented project is to provide useful insights on the budding MobileCoin network by crawling the nodes participating in consensus.
The results of this project can be summarised as follows:
\begin{itemize}
    \item It is possible to accurately crawl the core MobileCoin network,
    obtaining a snapshot of the quorum system induced by individual configurations---the MobileCoin \acs{fbas} (Section \ref{sec:crawl}).
    \item An off-the-shelf laptop with Intel SGX support will likely not meet the requirements to operate a validator node (Section \ref{sec:crawl}).
    \item Safety and liveness guarantees are currently determined by the quorum set configurations of 10 nodes distributed across 7 organisations (Section \ref{sec:fbas}).
    \item The MobileCoin quorum system prioritises safety at the expense of a higher risk of loss of liveness (see Section \ref{sec:fbas}).
\end{itemize}
To the best of our knowledge,
we are first to study the MobileCoin \acs{fbas}
and publicly present our findings.
Complementary to this report,
we maintain an interactive online version of our analyses and make our collected crawl data available\footnote{\url{https://trudi.weizenbaum-institut.de/mobilecoin_analysis/}}.

\section{Federated Byzantine Agreement Systems}\label{sec:fbas_basics}

Being based on \ac{mcp}, MobileCoin induces a \ac{fbas} \cite{mazieres2015stellar}.
The structure of the MobileCoin \ac{fbas} has a determining effect on key security and performance properties of the MobileCoin payments network.
In the following, we give an informal overview of key concepts and terminology related to \acp{fbas}.
We defer to works such as \cite{florian2021sum} for a more in-depth exploration of the topic.

An \ac{fbas} constitutes of a set of nodes, each of which is associated with an individual description of which groups of other nodes it requires agreement from.
\ac{mcp} implements this notion using the concept of \textit{quorum sets}.
A quorum set defines a group of nodes and optional inner quorum sets along with a threshold value denoting how many of the quorum set's constituents must be in agreement.

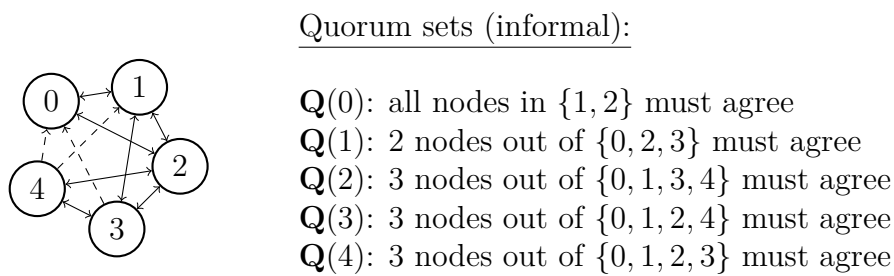
\begin{figure}[htpb]
  \centering
    \begin{minipage}{0.3\textwidth}
      \begin{center}
          \begin{tikzpicture}[
  dot/.style = {circle, inner sep=0pt, outer sep=0pt}
  ]
  \def \topradius {1cm}
  \def \midradius {2cm}
  \def \botradius {3cm}

  \foreach \s in {0,...,4}
  {
    \node[draw, circle, thick] (v\s) at ({135 - 360/5 * \s}:\topradius) {$\s$};
  }
  \path [<->] (v0) edge [draw] (v1);
  \path [<->] (v0) edge [draw] (v2);
  \path [->, dashed]  (v3) edge [draw] (v0);
  \path [->, dashed]  (v4) edge [draw] (v0);

  \path [<->] (v1) edge [draw] (v2);
  \path [<->] (v1) edge [draw] (v3);
  \path [->, dashed]  (v4) edge [draw] (v1);

  \path [<->] (v2) edge [draw] (v3);
  \path [<->] (v2) edge [draw] (v4);

  \path [<->] (v3) edge [draw] (v4);

\end{tikzpicture}
      \end{center}
    \end{minipage}
    \begin{minipage}{0.6\textwidth}
      \underline{Quorum sets (informal):}\\
      \\
      $\Q(0)$: all nodes in $\{1,2\}$ must agree\\
      $\Q(1)$: 2 nodes out of $\{0,2,3\}$ must agree\\
      $\Q(2)$: 3 nodes out of $\{0,1,3,4\}$ must agree\\
      $\Q(3)$: 3 nodes out of $\{0,1,2,4\}$ must agree\\
      $\Q(4)$: 3 nodes out of $\{0,1,2,3\}$ must agree\\
    \end{minipage}
    \caption{An example \ac{fbas} $(\V, \Q)$ with nodes $\V = \set{0,1,2,3,4}$
      and quorum sets $\Q$ (on the right) defined in an informal manner.
      Also depicted (on the left) is a heuristic graph representation of the \ac{fbas}.
    }%
  \label{fig:fbas_example}
\end{figure}

Figure \ref{fig:fbas_example} illustrates this abstract definition with an example \ac{fbas}.
Depicted are the quorum sets of five nodes (on the right; there are no inner quorum sets)
and a heuristic graph representation of the resulting \ac{fbas}.
Note that any representation of an \ac{fbas} as a regular graph is necessarily incomplete,
as constraints such as \enquote{$m$ out of $n$ nodes must agree} cannot be modelled.
In the example, node $0$ requires agreement from both nodes $1$ and $2$, whereas node $1$ requires agreement from at least two nodes out of the set $\{0, 2, 3\}$.
At least one of the combinations $\{0, 2\}$, $\{0, 3\}$, $\{2, 3\}$ must agree on the same (valid) value in order for node $1$ to accept and eventually confirm that value.
In MobileCoin, the \emph{value} typically corresponds to the contents of the next blockchain block.

Informally, an \ac{fbas} enables \textit{liveness} for a consensus protocol like \ac{mcp} if,
when honouring the configurations of all nodes,
the consensus protocol is able to make progress,
e.g. extend a blockchain with new blocks.
\ac{fbas} nodes agree on new values only when sufficient nodes in their quorum sets agree on the same value.
In the above example, node $0$ will only accept a new block if it is certain than nodes $1$ and $2$ will accept it as well.
A group of \ac{fbas} nodes that can
by itself agree on new values is called a \emph{quorum}.
In the above example, nodes $\set{0,1,2,3}$ form a quorum, but nodes $\set{0,1,2}$ do not.
Sets of nodes that intersect every quorum in an \ac{fbas} are known as \textit{blocking sets}.
If all nodes in a blocking set crash or become uncooperative, no quorums can be formed and liveness is necessarily lost.
In the above example, $\set{2}$ is already a blocking set.

In order to guarantee \textit{safety},
i.e. that no two sets of nodes agree on conflicting values,
an \ac{fbas} must enjoy \textit{quorum intersection}.
This means that each two quorums should share at least one common node.
Lack of quorum intersection in an \ac{fbas} can, for example, lead to forks and double spends.
A \textit{splitting set} is a set of nodes that contains an intersection of at least two quorums.
If a splitting set consists of malicious nodes, these nodes can conspire to agree to different statements in two quorums they are \enquote{splitting}, causing divergence and inconsistency.
In the above example, $\set{0,1,2}$ form a splitting set.

An important part of \ac{fbas} analysis \cite{florian2021sum} consists in determining minimal blocking sets and minimal splitting sets,
thereby giving a lower bound for the number of nodes that need to be compromised in order to threaten liveness or safety.

\section{Crawling the MobileCoin Network}\label{sec:crawl}

In the following, we discuss approaches to crawling the MobileCoin network so as to gather empirical data on the network's health and the structure of the MobileCoin \ac{fbas}.

\subsection{Network Structure}

There are three types of nodes in the MobileCoin network: (1) (full) validators, (2) watchers and (3) clients.
Validator nodes actively participate in consensus, validate transactions and potentially advance the ledger---depending on their role in other nodes' quorum sets.
In essence, validators are comparable to full nodes in Bitcoin and similar Proof-of-Work blockchains.
Watchers on the other hand play an observing role and verify the validity of the ledger produced by validators.
Regarding consensus, watcher nodes are entirely passive.
Clients, such as the Signal messaging app,
obtain ledger data from individual watchers with local synchronised copies of the blockchain,
trusting them to determine the correct state of the ledger.

Only validator nodes therefore participate in consensus, i.e. \ac{scp}.
In order to become a validator node and participate in consensus,
a machine has to meet the hardware requirements published by MobileCoin%
\footnote{\url{https://github.com/mobilecoinfoundation/mobilecoin/blob/master/consensus/service/BUILD.md}}%
.
Most notably, an Intel SGX-enabled processor is a necessity which rules out many popular cloud providers and off-the-shelf equipment.

Notwithstanding, communication with validators in the network is also possible for non-validator nodes.
Validators provide a public RPC API enabling queries for consensus protocol-related information.
With this API it is possible to extract a validator node's quorum set configuration
(a \enquote{piece of the MobileCoin \ac{fbas}}, so to speak),
as well as the network-layer addresses under which the nodes referenced in a node's quorum set can be reached.
The latter is due to the fact that MobileCoin node operators are required to specify which nodes their nodes should connect to, as there is no peer discovery or similar networking logic.
Hence, if reasonably configured, each node will have connections to at least the members of its quorum set because connections to these nodes are needed for consensus.

\subsection{Crawler and Crawl Procedure}
\label{sub:crawler}

We implemented a crawler leveraging the aforementioned public API of MobileCoin validator nodes\footnote{\url{https://github.com/wiberlin/mc-crawler}}.
Our crawler operates at the application layer of the MobileCoin network and collects data about each validator such as its host name, public key and quorum set.

Our crawler queries validators using the \code{GetLatestMsg} RPC.
\code{GetLatestMsg} asks a validator for the last consensus message it broadcast to the other validators.
The response to the RPC is a message type that is exchanged between nodes during consensus\footnote{\url{https://github.com/mobilecoinfoundation/mobilecoin/blob/master/peers/src/consensus\_msg.rs\#L20}}.
It includes an \ac{scp} message, the ID of the block the message should be appended to as well as the \ac{scp} message signature.
The \ac{scp} message in the consensus message consists of the sending node's ID, the slot index, the queried node's quorum set and the payload\footnote{\url{https://github.com/mobilecoinfoundation/mobilecoin/blob/master/consensus/scp/src/msg.rs\#L276}}.
The crawler extracts the queried node's quorum set from the \ac{scp} message which, for the sake of crawling the network, is seen as a list of nodes.
The crawler iterates through known nodes, sending the RPC to each node it has not yet communicated with and updating the list of nodes-to-be-contacted accordingly.
The crawler terminates when its list of nodes to be crawled is empty, i.e. no quorum set in any RPC response contained a new node.
We bootstrap the crawl from a list of validator node addresses published on the MobileCoin Foundation's website.

\subsection{Potential and Limitation of the Crawl Methodology}
\label{sub:limitations}

Implementing the crawler as a client has the benefit of low-resource usage and straight-forward deployment, as Intel SGX support is not necessary.
However, our approach also has two limitations:
\begin{itemize}
    \item Watcher nodes, i.e. nodes whose main task is to verify the signatures attached to blocks by full validators, cannot be queried using our method.
    \item Validators that are not included in at least one quorum set will not be found.
\end{itemize}

Watcher nodes are inherently hard to systematically monitor, as these nodes do not participate in consensus and there is no peer discovery in the MobileCoin network which could enable their enumeration.
A potential angle for analyses is to run two or more validators and to infer properties of the watcher population from the sample of watchers that connect to these watchers (cf. \cite{balduf2021monitoring}).
We leave these considerations for future work.

Regarding the second limitation: nodes that are not included in any crawled nodes' quorum sets are either not part of the same \ac{fbas},
or are guaranteed to be outside of the MobileCoin \ac{fbas}' \emph{top tier}~\cite{florian2021sum}.
In other words, the nodes our crawler cannot reach are essentially irrelevant when determining the safety and liveness properties of the network in which we start our crawl,
which
(since we start our crawl from nodes trusted by the MobileCoin foundation)
presumably is the \enquote{true} MobileCoin network.

Summarising, our method gives a sufficiently informative account of the network to assess consensus-related questions of interest.

\section{Analysis of the MobileCoin FBAS}\label{sec:fbas}

We now present the results of our crawls of the MobileCoin network.
We first discuss node-related data points,
followed by the presentation and analysis of the MobileCoin quorum structure as inferred from our crawl data.
The results discussed in the following were gathered between August 23$^{rd}$ and November 23$^{rd}$ 2021.
We started a new crawl on each full hour during this period.
Currently, a full crawl takes approximately 2 seconds.
Short crawl durations are important for ensuring the accuracy of obtained network snapshots~\cite{stutzbach2006capturing}.

\subsection{Nodes}
\label{sub:nodes}

The crawler consistently reported the same 10 nodes during each crawl.
The relatively low number of nodes can be expected given that our crawler discovers only core validator nodes
(cf. Section~\ref{sub:crawler})
and the public network was announced in December 2020\footnote{\url{https://mobilecoinfoundation.medium.com/mobilecoin-main-net-8e355d82c726}}.
During the vast majority of our hourly crawls,
100\% of the nodes were online, i.e. the crawler got a response to every RPC it sent
(see the number of nodes reported in Figure~\ref{fig:sets_in_time}).

By manually examining each node's host name and network-layer address,
we were able to infer a name for the organisation operating each node,
as well as the country\footnote{Data collected from the Geolite2 databases accessible at \url{https://www.maxmind.com}.} in which the node is located and the ISP\footnote{Data collected using \url{https://ipinfo.io}.} it is hosted at.
As shown in Table \ref{tab:orgs}, there are currently 7 different active organisations in the network.
Tables \ref{tab:country} and \ref{tab:isp} illustrate the distribution of the 10 nodes across countries and ISPs respectively.
90\% of the nodes are hosted at Microsoft Corporation while 70\% of the nodes are located in the Netherlands.
Both forms of centralisation can be attributed to the limited availability of cloud providers offering machines with Intel SGX support.
Potential effects this centralisation may have on the robustness of the network will be discussed in the following (Section~\ref{sub:analysis}).
\begin{table}[ht]
    \caption{Node distribution over various entities.}
    \label{ab:groups}
    \begin{minipage}[t!]{.5\linewidth}
        \centering
        \begin{subtable}{\textwidth}
            \centering
            \caption{Distribution over Organisations}
            \label{tab:orgs}
            \begin{tabular}{ |p{7em}|c| }
                \hline
                Organisation & Node Count\\
                \hline
                MobileCoin\newline Worldwide & 3\\
                \hline
                Namda & 2 \\
                \hline
                Binance & 1 \\
                \hline
                Blockdaemon & 1 \\
                \hline
                Dreamhost & 1 \\
                \hline
                Ideas Beyond\newline Borders & 1 \\
                \hline
                The Long Now Foundation & 1 \\
                \hline
            \end{tabular}
        \end{subtable}
    \end{minipage}
    \begin{minipage}{.5\linewidth}
        \centering
        \begin{subtable}{\textwidth}
            \centering
            \caption{Distribution over Countries}
            \label{tab:country}
            \begin{tabular}{ |c|c| }
                \hline
                Country & Node Count\\
                \hline
                NL & 7\\
                \hline
                GB & 2 \\
                \hline
                DE & 1 \\
                \hline
            \end{tabular}
        \end{subtable}        
        \bigskip
        \begin{subtable}{.7\linewidth}
            \centering
            \caption{Distribution over ISPs}
            \label{tab:isp}
            \begin{tabular}{ |c|c| }
                \hline
                ISP & Node Count\\
                \hline
                Microsoft Corporation & 9\\
                \hline
                Datacamp Limited & 1\\
                \hline
            \end{tabular}
        \end{subtable}
    \end{minipage}%
\end{table}%

\subsection{Quorum Structure}
\label{sub:analysis}

The MobileCoin \ac{fbas} currently comprises of the 10 nodes discussed in Section~\ref{sub:nodes},
with each node referencing the other 9 nodes in its quorum set\footnote{
  Unlike nodes in the Stellar network \cite{lokhava2019stellar_payments},
MobileCoin node operators do not seem to follow the convention of including each node in its own quorum set.
}.
We illustrate the crawled quorum set configurations in Figure \ref{fig:fbas_mobilecoin}.
The nodes in each node's quorum set are weighted equally,
with a quorum set threshold of 7.
In other words, each of the crawled MobileCoin nodes requires agreement from at least 7 of the other 9 nodes.
The crawled nodes form a \textit{symmetric top tier} as per \cite{florian2021sum}.
This implies that node operators likely coordinate their configurations with each other to some extent.
The \ac{fbas} clearly enjoys quorum intersection,
and hence provides the basic preconditions to safe consensus.

\begin{figure}[htpb]
  \centering
    \begin{minipage}{0.4\textwidth}
      \begin{center}
          \begin{tikzpicture}[transform shape]
    \def\n{10}
    \def\nodenames{{"MC1", "MC2", "MC3", "Na1", "Na2", "Bin", "Blo", "Dre", "IBB", "LNF"}}
    \node[circle,minimum size=5cm] (b) {};
    \pgfmathtruncatemacro\nmo{\n-1}
    \foreach\x in{0,...,\nmo}{
        \pgfmathsetmacro{\name}{\nodenames[\x]}
        \node[minimum size=0.25cm,draw,circle,thick] (n-\x) at (b.{135 - 360/\n*\x}){\tiny \name};
    }
    \pgfmathtruncatemacro\nmt{\n-2}
    \foreach\x in {0,...,\nmt}{
        \pgfmathtruncatemacro\xpo{\x+1}
        \foreach\y in {\xpo,...,\nmo}{
            \ifnum\x=\y\relax\else
            \draw (n-\x) edge[<->] (n-\y);
            \fi
        }
    }
\end{tikzpicture}
      \end{center}
    \end{minipage}
    \begin{minipage}{0.5\textwidth}
      \underline{Quorum sets (informal):}\\
      \\
      For each node $v \in \V$ ($\len{V} = 10$):\\
      $\Q(v)$: $7$ other nodes ($\in \V \setminus \set{v}$) must agree\\
      $\iff$ $8$ nodes (of all nodes $\V$) must agree
    \end{minipage}
    \caption{Informal description of the crawled MobileCoin \ac{fbas},
      with quorum sets $\Q$ (on the right)
      and a heuristic graph representation (on the left).
      Nodes in the graph are named based on the organisations operating them.
    }%
  \label{fig:fbas_mobilecoin}
\end{figure}

We further analyse the MobileCoin \ac{fbas} using the \code{fbas\_analyzer}\footnote{\url{https://github.com/wiberlin/fbas\_analyzer}} framework.
In the following, we consider \textit{minimal blocking sets} and \textit{minimal splitting sets} as defined in \cite{florian2021sum}.
These are minimal sets that are sufficient to compromise, respectively,
the liveness and safety of an \ac{fbas}
(see our short overview in Section \ref{sec:fbas_basics}).
We occasionally merge nodes together based on their organisation, country or ISP.
For example, when merged according to organisation, we say that there are 7 nodes in the MobileCoin network (see Table \ref{tab:orgs}).

\begin{figure}[htpb]
  \begin{subfigure}[t]{0.5\textwidth}
    \centering
    \includegraphics[width=1.1\textwidth]{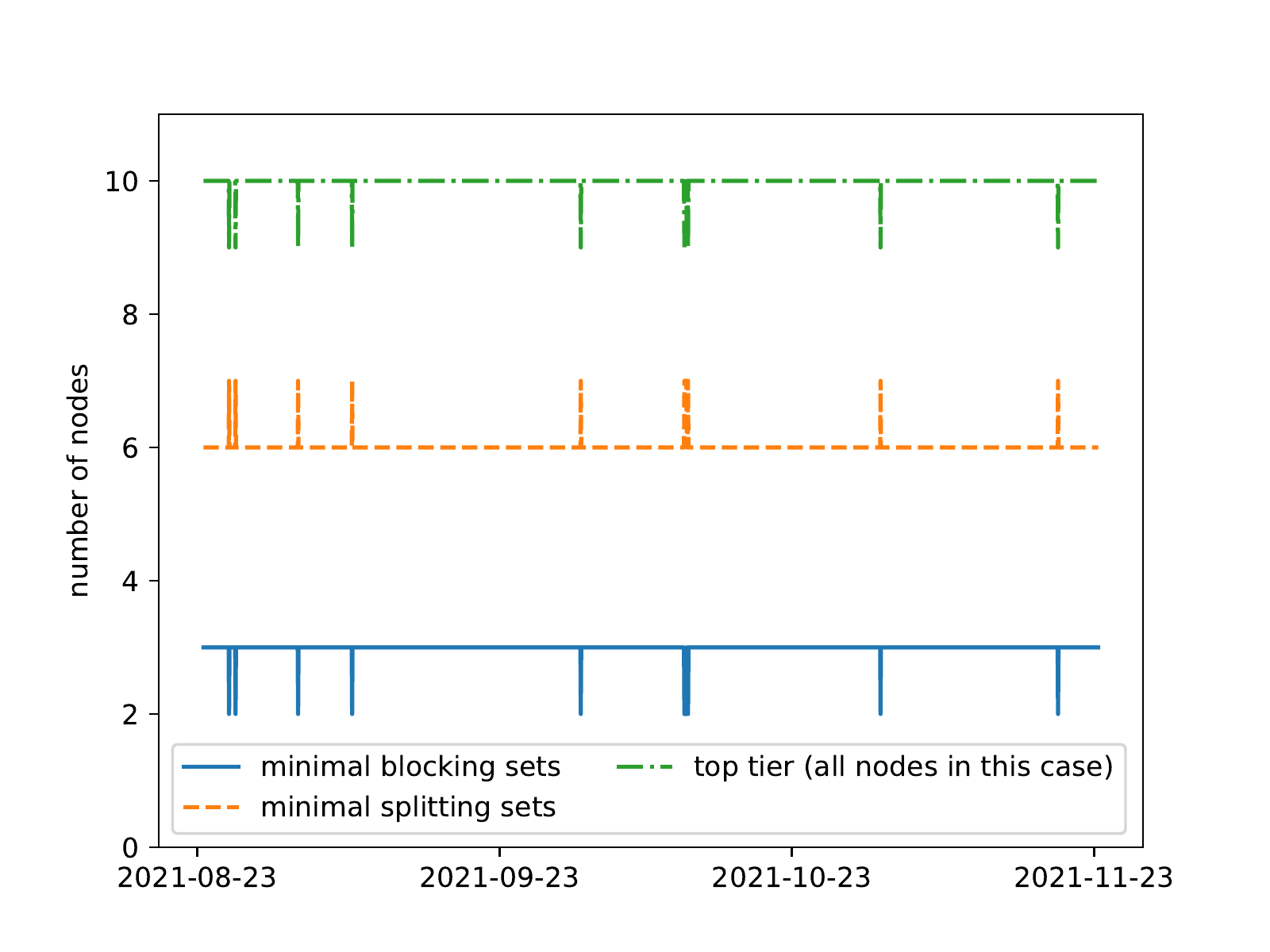}
    \caption{
      Results from all of our crawls so far (hourly).
      For each presented \ac{fbas} snapshot,
      the plot charts the size of its top tier as well as the mean cardinalities of minimal blocking and minimal splitting sets,
      with area boundaries marking the cardinalities of the smallest and largest respective set.
      All spikes are due to single nodes being unreachable during our crawl---%
      we assume that they have been inactive at that time.
    }
    \label{fig:sets_in_time}
  \end{subfigure}
  \begin{subfigure}[t]{0.4\textwidth}
    \centering
    \includegraphics[width=6.5cm, height=6.5cm]{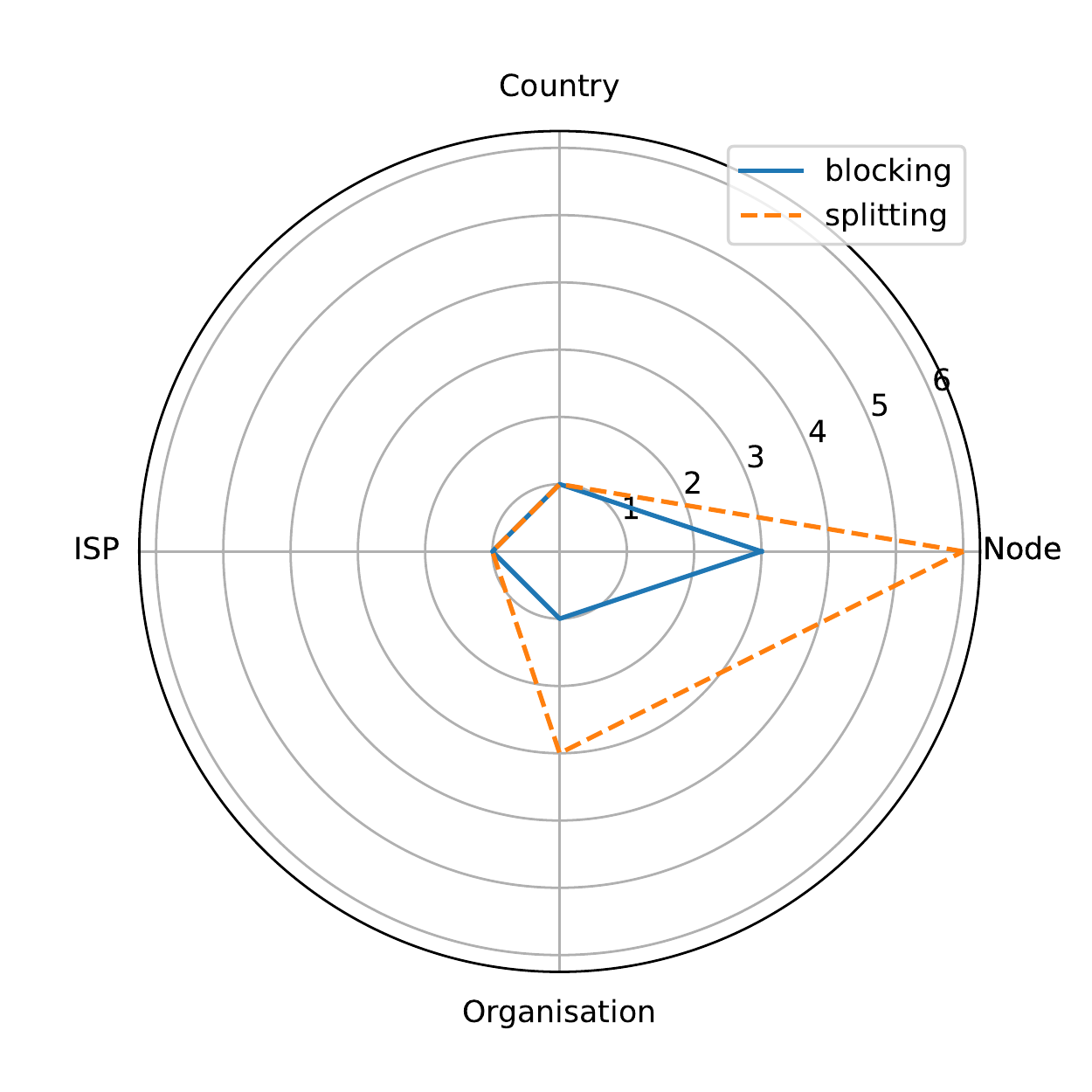}
    \caption{The size of the smallest blocking and splitting sets for \ac{fbas} snapshots in which all nodes were available.
      The northern, western and southern axes result from merging together nodes that are,
      respectively,
      located in the same country, hosted at the same ISP or operated by the same organisation.
    }
    \label{fig:sets_radar}
  \end{subfigure}
  \caption{Sets of nodes able to compromise safety and liveness in the MobileCoin \ac{fbas}, as per our crawls from August 2021 to November 2021.}
  \label{fig:analyzer_results}
\end{figure}

The results of our analyses are presented in Figure~\ref{fig:analyzer_results}.
Figure \ref{fig:sets_in_time} shows the sizes of the respective minimal sets as well as the size of the \ac{fbas}' top tier for all crawls performed in the discussed period.
In the analysed \ac{fbas} snapshots, each node is also a top tier node,
i.e. the top tier is equivalent to the set of all nodes.
Some nodes were not always reachable.
We mark nodes unreachable by our crawler as inactive,
resulting in a smaller \ac{fbas} with slightly different blocking and splitting sets.
Figure \ref{fig:sets_radar} depicts the sizes of the minimal blocking sets and minimal splitting sets for the case in which all nodes are available.
The figure shows how set size varies when nodes are merged by organisation, ISP or country.

Merged by organisations,
MobileCoin Worldwide
forms a minimal blocking set,
i.e. MobileCoin Worldwide could, in principle, stop the network from reaching consensus.
A possible improvement to this situation entails growing the \ac{fbas},
respectively its top tier,
with more validator nodes that are not controlled by this organisation.
Alternatively, the threshold values of current nodes' quorum sets could be adapted at the expense of increased safety risks.
Specifically,
reducing all quorum set thresholds by $1$ would result in both the smallest blocking sets and the smallest splitting sets being comprised by $4$ nodes in the control of $2$ organisations. %

Threats also emerge when considering the MobileCoin \ac{fbas} at the level of ISPs and countries.
For example, as an ISP, Microsoft Corporation by itself forms both a minimal blocking set and a minimal splitting set,
while the Netherlands on its own forms a minimal blocking set.
If Microsoft Corporation or the Netherlands forbid MobileCoin traffic or shut down all nodes within their area of influence,
the MobileCoin network would no longer be able to reach consensus until the operators of remaining nodes adapt their quorum set configurations.
In the highly unlikely case that Microsoft Corporation,
which is likely not only the ISP but also the hosting provider of 9 of the 10 MobileCoin validators,
decides to take control over the MobileCoin nodes in its area of influence,
it would even be able to split the network and induce forks.
To eliminate these threat scenarios,
node operators would need to increase the diversity of countries and ISPs (respectively hosting providers) that their nodes are hosted at.
This might be non-trivial in the case of hosting providers,
as currently only few hosting providers seem to offer (virtual) machines with sufficient Intel SGX capabilities.
In addition to making the deployment of new validator nodes more difficult,
the Intel SGX-related hardware requirements to MobileCoin nodes might therefore also lead to a centralisation in terms of ISPs and hosting providers.
Both of these issues might resolve themselves in the future with improvements in SGX availability.

\section{Conclusion}%
\label{sec:conclusion}
We present an application-layer crawler that is able to enumerate validators in the MobileCoin network.
Based on the data collected over a period of 3 months, we analyse the MobileCoin \ac{fbas} and report on our findings.
We find that the MobileCoin quorum system consistently constituted of the same 10 validators operated by 7 organisations over the observed duration.
Furthermore, while the validators' quorum set configurations undoubtedly induce quorum intersection and high levels of safety,
our analysis also uncovers several risks to liveness that should be addressed in the future.

MobileCoin uses Intel SGX enclaves to further improve the privacy of MobileCoin users.
However, the resulting hardware requirements for validators also make it harder to deploy validator nodes and limit the selection of hosting providers for such nodes.
Both of these consequences of MobileCoin's SGX usage could lead to centralisation effects.

We continue to crawl the network, with results published to our interactive analysis website\footnote{\url{https://trudi.weizenbaum-institut.de/mobilecoin_analysis/}}.
As a possible avenue for follow-up works,
a methodology could be developed for also enumerating the population of MobileCoin watcher nodes.
Additionally, various questions are still open w.r.t. to the FBAS paradigm itself,
such as whether and how FBASs can support the notion of fair per-block validator rewards.

\bibliography{bib}

\begin{thebibliography}{6}
\providecommand{\natexlab}[1]{#1}
\providecommand{\url}[1]{\texttt{#1}}
\expandafter\ifx\csname urlstyle\endcsname\relax
  \providecommand{\doi}[1]{doi: #1}\else
  \providecommand{\doi}{doi: \begingroup \urlstyle{rm}\Url}\fi

\bibitem[mob(2017)]{mobilecoin2017}
{MobileCoin}.
\newblock \url{https://buymobilecoin.com/assets/MobileCoin_White_Paper.pdf}, 11
  2017.
\newblock accessed: August 2021.

\bibitem[Mazi\`{e}res(2015)]{mazieres2015stellar}
David Mazi\`{e}res.
\newblock The {Stellar} consensus protocol: A federated model for
  internet-level consensus, 2015.
\newblock URL \url{https://stellar.org/papers/stellar-consensus-protocol.pdf}.

\bibitem[Lokhava et~al.(2019)Lokhava, Losa, Mazi\`{e}res, Hoare, Barry, Gafni,
  Jove, Malinowsky, and McCaleb]{lokhava2019stellar_payments}
Marta Lokhava, Giuliano Losa, David Mazi\`{e}res, Graydon Hoare, Nicolas Barry,
  Eli Gafni, Jonathan Jove, Rafa\l{} Malinowsky, and Jed McCaleb.
\newblock Fast and secure global payments with {Stellar}.
\newblock In \emph{Proceedings of the 27th ACM Symposium on Operating Systems
  Principles}, page 80–96, New York, NY, USA, 2019. Association for Computing
  Machinery.
\newblock ISBN 9781450368735.

\bibitem[Florian et~al.(2021)Florian, Henningsen, Ndolo, and
  Scheuermann]{florian2021sum}
Martin Florian, Sebastian Henningsen, Charmaine Ndolo, and Björn Scheuermann.
\newblock The sum of its parts: Analysis of federated byzantine agreement
  systems.
\newblock \url{https://arxiv.org/abs/2002.08101}, 2021.

\bibitem[Balduf et~al.(2021)Balduf, Henningsen, Florian, Rust, and
  Scheuermann]{balduf2021monitoring}
Leonhard Balduf, Sebastian Henningsen, Martin Florian, Sebastian Rust, and
  Björn Scheuermann.
\newblock Monitoring data requests in decentralized data storage systems: A
  case study of {IPFS}.
\newblock \url{https://arxiv.org/abs/2104.09202}, 2021.

\bibitem[Stutzbach and Rejaie(2006)]{stutzbach2006capturing}
Daniel Stutzbach and Reza Rejaie.
\newblock Capturing accurate snapshots of the {G}nutella network.
\newblock In \emph{Proceedings IEEE INFOCOM 2006. 25TH IEEE International
  Conference on Computer Communications}, pages 1--6, 2006.
\newblock \doi{10.1109/INFOCOM.2006.347}.

\end{thebibliography}

\end{document}